\documentclass[aps, prb, twocolumn, superscriptaddress]{revtex4-1}
\usepackage{hyperref}
\usepackage{amsmath}
\usepackage{amsfonts}
\usepackage{amssymb}
\usepackage{graphicx}
\usepackage{color}

\begin{document}

\title{Efficient Dynamic Polarization of Phosphorus Nuclei in Silicon in Strong Magnetic Fields and Low Temperatures}
\date{\today}
\author{J. J\"{a}rvinen}
\email{jaanja@utu.fi}
\author{D. Zvezdov}
\altaffiliation{Institute of Physics, Kazan Federal University, Russia}
\author{J. Ahokas}
\author{S. Sheludyakov}
\author{O. Vainio}
\author{L. Lehtonen}
\author{S. Vasiliev}
\affiliation{Wihuri Physical Laboratory, Department of Physics and Astronomy, University of Turku, 20014 Turku, Finland}
\author{Y. Fujii}
\author{S. Mitsudo}
\author{T. Mizusaki}
\affiliation{Research Center for Development of Far-Infrared Region, University of Fukui, 3-9-1 Bunkyo, Fukui 910-8507, Japan}
\author{M. Gwak}
\author{SangGap Lee}
\affiliation{Division of Materials Science, Korea Basic Science Institute, 169-148 Gwahak-ro, Yuseong-gu, Daejeon 305-806, Korea}
\author{Soonchil Lee}
\affiliation{Department of Physics, Korea Advanced Institute of Science and Technology, 291 Daehak-ro, Yuseong-gu, Daejeon 305-701, Korea}
\author{L. Vlasenko}
\affiliation{A. F. Ioffe Physico-Technical Institute, Russian Academy of Sciences, 194021 St. Petersburg, Russia}


\begin{abstract}
We demonstrate that the dynamic nuclear polarization (DNP) of phosphorus donors in silicon can be very effective in a magnetic field of 4.6 T and at temperatures below 1 K. The DNP occurs due to the Overhauser effect following a cross relaxation via the forbidden flip-flop or
flip-flip transitions. Nuclear polarization values $P>0.98$ were reached after 20 min of pumping with 0.4 $\mu$W of microwave power. We evaluated
that the ratio of hyperfine state populations increased by three orders of magnitude after 2 hours of pumping, and an extremely pure nuclear spin system containing $<10^{-11}$ of the other spin state can be created. An inverted DNP has been obtained by pumping the low field
ESR line of P followed by the flip-flip cross relaxation. This transition has much smaller relaxation rate and required substantially longer
pumping times. We found that the nuclear polarization dynamics deviates substantially from a simple exponential function. The evolution of the
polarization is characterized by two time constants $T^{'}_{ac}\approx$20 s in the beginning, and $T^{''}_{ac}\approx$1100 s for long pumping
time. Temperature dependence of the nuclear relaxation rate of $^{31}$P was studied down to 0.75 K, below which the relaxation time became too
long to be measured. The nuclear polarization followed a bi-exponential time dependence during relaxation. We suggest that the non-exponential
behavior of DNP dynamics and the subsequent relaxation is mediated by the nuclei of $^{29}$Si surrounding $^{31}$P donors, which affect the
transition probabilities of the forbidden cross-relaxation processes.

\end{abstract}
\maketitle

\section{Introduction}
Shallow donors P, As and Bi in silicon have been studied extensively since the pioneering experiments of Feher \cite{Feher}. A revived interest
to these systems has been raised by a recent proposal of using donor spins as qubits for quantum computing \cite{Kane}. A long coherence time
and ease of qubit initialization by external microwave fields are necessary conditions for building a quantum computer. These properties, also
known as long transversal relaxation time and fast dynamic nuclear polarization, are inherent features of P impurities in Si (Si:P) at cryogenic
temperatures. In this work we will concern the dynamic nuclear polarization and relaxation of P donors in silicon at low temperatures and high
magnetic fields, when their electron spins are polarized to a very high degree, while the nuclear spins are not yet polarized.

The Overhauser effect (OE) is one of the well known and most effective ways of DNP. In the OE the allowed electron spin resonance transitions
are saturated, and subsequent cross relaxation with the simultaneous spin flips of electron and nucleus leads to the population transfer from
one nuclear state to another. The rate of the OE DNP is determined by the cross relaxation. There are two possibilities for this process: with
an opposite change of the electron and nuclear spins, known as flip-flop transitions; and with flip-flip transitions, when the electron and
nuclear spins flip to the same direction. Both transition probabilities strongly depend on the degree of anisotropy of the interaction between
electron and nucleus. Dipole-dipole interaction of the host lattice nuclei with donor electrons has large anisotropy and ensures fast relaxation
rate. Due to this reason the OE DNP has been successfully utilized for polarization of the nuclei in solids and liquids via the embedded
paramagnetic ions (see \cite{Maly2008, Atsarkin2011} for recent reviews). Situation is different for the isolated ion or donor atom with
unpaired $s$-electron. Interaction of the electron with its own (core) nuclei is of isotropic Fermi type, and the flip-flop transition
probability is given by the hyperfine mixing factor $\eta=A/2\hbar \gamma_{e} B_{0}$, with $A$- being the hyperfine constant, $\gamma_e$ - the
electron gyromagnetic ratio, and $B_0$ - applied static magnetic field. The flip-flip transition is fully forbidden in this case. Therefore,
efficiency of the OE effect for polarizing donor nuclei was considered to be low, especially in high magnetic fields \cite{Maly2008, vanTol}.

A second important factor for effective DNP is the nuclear relaxation rate, which is the process responsible for returning the polarization to thermal equilibrium and thus countering the effect of DNP. Main
relaxation mechanism at low temperatures is the nuclear Orbach process, which also proceed via the flip-flop and flip-flip transitions with the
rate weighted by the Boltzmann factor \cite{AbragamGoldman}. At low temperatures the nuclear relaxation rate decreases exponentially, increasing
efficiency of OE DNP.

Despite of the numerous studies, Si:P system has not been thoroughly investigated in strong magnetic fields and we are not aware of any works
below 1 K. The highest DNP value of $P> 0.75$ has been reached in 8.6 T magnetic field \cite{vanTol} and at temperatures of $\approx3$ K. The
nuclear relaxation rate turned out to be fast in these conditions which, regardless of the high electron spin polarization ($>98\%$), limited
reaching a higher nuclear polarization. Typically in DNP experiments relatively high values of microwave power (up to 1 W) were used for
irradiating Si:P samples.

In the present work we demonstrate that an efficiency of OE DNP is very high in strong magnetic fields and low temperatures. We cooled Si:P
samples to $\approx0.2$ K, which led to a decrease of the nuclear relaxation rate to immeasurably small values and allowed reaching nuclear
polarization $P\geq0.98$ after excitation of 20 minutes. We suggest that extremely pure nuclear spin states with $1-P < 10^{-11}$ can be
produced with sub-$\mu$W power levels. We evaluated the cross relaxation rate by measuring the rate of the DNP, in which the flip-flop
transition is the limiting step.

\section{Experimental details}
\begin{figure}
\begin{center}
\includegraphics[%
  width=0.95\linewidth,
  keepaspectratio]{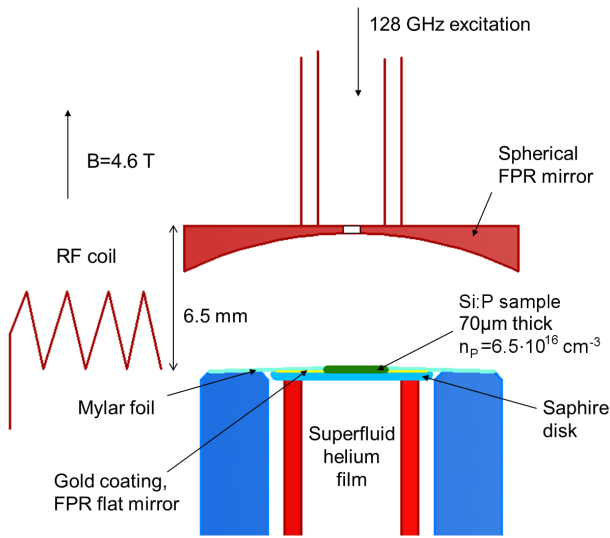}
\end{center}
\caption{(Color online) Scheme of the experimental setup with Fabry-Perot resonator.} \label{ExpCell}
\end{figure}
The experimental cell (SC) with the Si:P sample was placed in the center of a superconductive magnet producing a field of 4.6 T and cooled down
by a dilution refrigerator (Fig. \ref{ExpCell}). The sample of crystalline silicon with natural abundance, 2$\times$2$\times$0.07 mm in size and
doped by 6.5$\times$10$^{16}$ cm$^{-3}$ of P, was placed onto the flat mirror of a Fabry-Perot resonator (FPR) having a $Q$ of approximately
4000. The crystal's [111] axis was directed along the axis of the FPR resonator and polarizing magnetic field. ESR spectra from the sample were
detected with a cryogenic heterodyne spectrometer \cite{Spectrometer} operating at 128 GHz. The spectrometer provides both absorption and
dispersion signal components without a modulation of the magnetic field and is optimized for reaching high sensitivity at a small excitation
power. This is important when studying samples with long electron spin-lattice relaxation times ($T_{1e}\approx 0.2$ s in our case). To avoid
the saturation of Si:P ESR lines below 1 K we had to use a very low microwave power for \textit{detection}, typically below 1 $p$W. For pumping
the ESR lines in the DNP experiments we used the maximum available microwave power of 0.4 $\mu$W, which will be further referred to as
\textit{pumping} power.

Cooling of the sample to 100 mK could be problematic because of the very large Kapitza resistance between the sample and the metal surface.
Thermal contact of the sample was improved with a superfluid helium. The flat mirror with the sample was isolated from the main vacuum by a
Mylar foil to a separate chamber, which could be filled with liquid helium (Fig. \ref{ExpCell}). However, in this work, we did not observe any
significant difference in the thermalization of Si:P samples with or without helium film at temperatures above 500 mK.

\section{Results}
\label{sec:Results}
\begin{figure}
\begin{center}
\includegraphics[%
  width=0.95\linewidth,
  keepaspectratio]{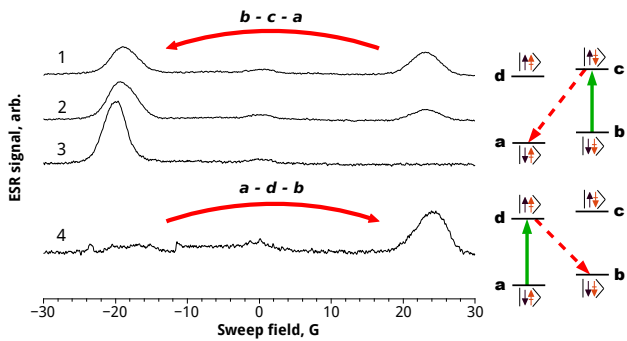}
\end{center}
\caption{(Color online) The ESR absorption spectra demonstrating DNP of P in silicon. Traces 1, 2 and 3 demonstrate the Overhauser effect:
pumping the \textit{b-c} transition, followed by the forbidden \textit{c-a} relaxation. Trace 1 - before pumping, trace 2 - after 100 s pumping
and trace 3 - after 20 min pumping. Trace 4 - demonstrates the result of pumping \textit{a-d} transition with the relaxation via the
\textit{d-b} transition. The transfer of hyperfine level populations is shown in the level diagram on the right. Solid arrows denote the allowed
ESR transitions and dashed traces relaxations via the forbidden flip-flop (upper diagram) and flip-flip (lower diagram) transitions. }
\label{DNPSpectra}
\end{figure}

The CW ESR absorption spectrum of a Si:P sample recorded just after cooling down to a low temperature is shown in the upper trace of Fig.
\ref{DNPSpectra}. The spectrum consists of two lines corresponding to the allowed \textit{a-d} (low field) and \textit{b-c} (high field)
transitions, separated by $\approx 42$ G due to the hyperfine interaction of $^{31}$P electron with its own nucleus. The lines are
inhomogeneously broadened by the spin-$1/2$ $^{29}$Si nuclei of the host lattice which reside inside the relatively disperse electron cloud of $^{31}$P
electron \cite{Feher}.

First, we performed a DNP experiment by saturating the allowed \textit{b-c} transition.  The magnetic field was set to the center of the
\textit{b-c} line  and the excitation power was increased to the pumping value, aiming at full saturation of the ESR transition. For increasing
the efficiency of the DNP the ESR excitation frequency was modulated with the frequency deviation corresponding to $3-4$ line widths and with
the modulation rate of $10-20$ Hz. Since the electron spin-lattice relaxation time $\approx$ 0.2 s is substantially longer than the modulation
period, all the spin packets in the ESR line are simultaneously saturated. After pumping the \textit{b-c} transition for a time $t_p$, the
spectrometer was switched to detection power, and undistorted spectra with both ESR lines were recorded. Then, the nuclear polarization
$P(t)=(n_a-n_b)/(n_a+n_b)$ was calculated from the ESR absorption line areas, which are proportional to the populations $n_a$ and $n_b$ of the
\textit{a} and \textit{b} states. The polarization measurement was performed shortly ($<50$ s) after the pumping, so that the nuclear state
populations were not influenced by the nuclear relaxation, which turned out to be very slow below 1 K.

\begin{figure}
\begin{center}
\includegraphics[%
  width=0.95\linewidth,
  keepaspectratio]{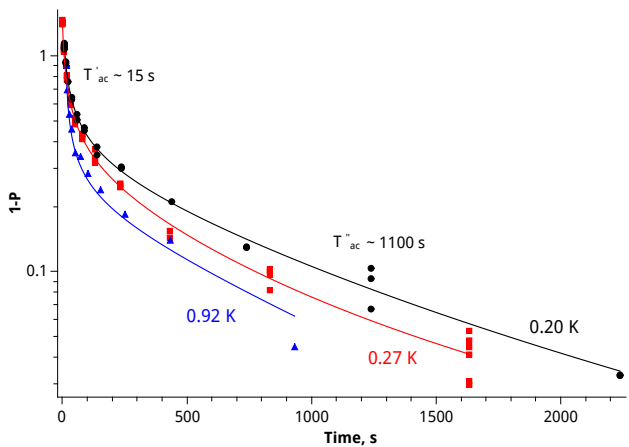}
\end{center}
\caption{(Color online) Dynamics of the DNP process via the Overhauser effect when pumping the \textit{b-c} transition. Log $1-P(t)$ is plotted as a function
of pumping time $t_p$ for 0.20 K - \textbf{\textbullet}, 0.27 K - {\color{red} $\blacksquare$ } and 0.92 K - {\color{blue} $\blacktriangle$}. The lines are guides for the eye.}\label{PolDynamics}
\end{figure}
The spectra after $t_p=30$ s and $t_p=20$ min are shown in the 2nd and 3rd traces of Fig. \ref{DNPSpectra}. The DNP occurs due to the
Overhauser effect, which transfers the population as: \textit{b}$\rightarrow$\textit{c}$\rightarrow$\textit{a} (the dashed arrow in Fig.
\ref{DNPSpectra} inset).

Due to the high $Q$ of the FPR and the long spin-lattice relaxation time of the $^{31}$P electrons, the allowed \emph{b-c} transition is fully
saturated in a fraction of a second. It is clear that the DNP rate is limited by the much slower relaxation via the forbidden \textit{a-c}
transition, characterized by a cross-relaxation time $T_{ac}$. As we shall see below, the nuclear polarization can be negative if $n_a<n_b$.
Therefore, to check whether the DNP behaves exponentially, we plotted $(1-P(t_p))$ in log units in Fig. \ref{PolDynamics}. One can see that the
polarization dynamics deviates strongly from an exponential at the beginning of the pumping. We verified that the ESR excitation power is strong
enough to saturate the allowed transitions. Pumping dynamics did not depend on excitation power near its maximum value. We have not observed any
reduction in the DNP efficiency after decreasing the power by 2-3 db. Only attenuating it by more than 10 db led to a decrease of the DNP. The
characteristic time of the DNP build-up is $T^{'}_{ac}\approx$20 s at the beginning and $T^{''}_{ac}\approx$1100 s at the end, when the function
is almost exponential. We estimate that the polarization $P>0.98$ was reached in this experiment, based on the signal-to-noise ratio of
$\approx$ 100 since the \textit{b-c} ESR line vanishes in the noise after $\approx$ 20 min of pumping.

Next, we pumped the \textit{a-d} transition. In this case the Overhauser effect leads to a DNP resulting in the population transfer
\textit{a}$\rightarrow$\textit{d}$\rightarrow$\textit{b} and the creation of negative polarization where $n_a<n_b$. The result  of this
experiment is shown in the trace 4 of Fig. \ref{DNPSpectra}. The rate of the process is substantially slower than the other cross relaxation
\textit{b}$\rightarrow$\textit{c}$\rightarrow$\textit{a}. The polarization was increased by a factor of 2 after about 10 h pumping. Since the
DNP in this case is limited by the flip-flip relaxation, we estimated the corresponding relaxation time $T_{bd}$ to be $\approx5\times10^4$ s.
This is  $\gtrsim 50$ times larger than $T^{''}_{ac}$.

We repeated the pumping experiment at temperatures 0.2 K, 0.27 K and 0.92 K and found out that the pumping dynamics follows a similar looking non-exponential curve. The DNP seems to proceed faster when the temperature increases from 0.2 to 0.9 K but above 1 K the nuclear relaxation is so fast that it starts to dominate the level populations. This indicates that the temperature dependence of the relaxation rate via the forbidden transitions is much weaker than the nuclear relaxation below 1 K. Another possible pathway for the DNP would be via the nuclear relaxation between the upper hyperfine states: $c\leftrightarrow d$. In this case, the slowest process, \textit{c-d} nuclear relaxation, will mediate the transfer rate. However, as we shall
see below, the nuclear relaxation rate has a very strong temperature dependence, which contradicts with the observation that $T_{ac}$ is only weakly temperature dependent.

\begin{figure}
\begin{center}
\includegraphics[%
  width=0.95\linewidth,
  keepaspectratio]{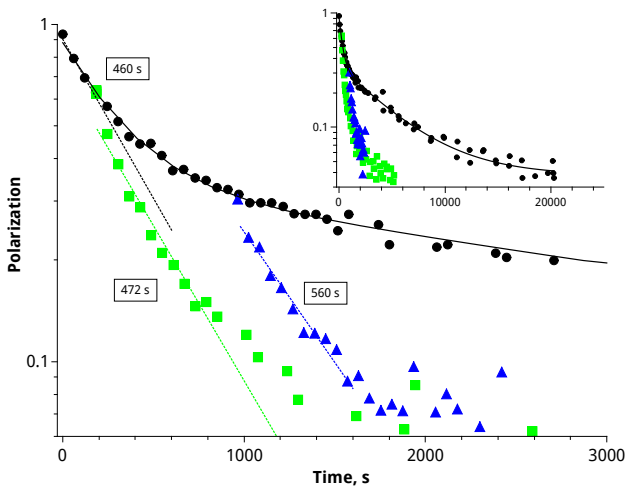}
\end{center}
\caption{(Color online) Nuclear relaxation measured with different starting polarization values at 1.37 K. \textbf{\textbullet} - relaxation curve after starting with $P_{0}\approx 1$, {\color{green}$\blacksquare$} and {\color{blue}$\blacktriangle$} - relaxation curves with lower
starting values of $P_{0}\approx 0.6$ and $P_{0}\approx 0.3$. The solid line is bi-exponential fit to the data. The dashed lines are exponential fits to the initial parts of the data. The data for $P_{0}\approx 0.6$ and
$P_{0}\approx 0.3$ are shifted in time to match the starting points with the $P_{0}\approx 1$ curve. The inset shows full time range of the relaxations.} \label{Rel2}
\end{figure}

At temperatures above 1 K, weak signals from the \textit{b-c} line were observed even after pumping it for a very long time  $t_p \gg
T^{''}_{ac}$. Since the nuclear \textit{a-b} relaxation is a mechanism competing with DNP, this process can be responsible for the decrease of
the maximum DNP value observed at higher temperatures. To verify this, the nuclear \emph{a-b} relaxation was studied of as a function of
temperature. We prepared starting conditions where all of the spins were in the \emph{a}-state by pumping the \textit{b-c} ESR line as described
above. The temperature was kept at the lowest value of 0.2 K, where the nuclear polarization does not relax within several days. Then, the
sample cell was rapidly warmed up and the temperature was stabilized to a desired value between 0.75 and 2.2 K. The evolution of the spin states
towards the thermal  equilibrium was monitored by measuring repeatedly the ESR spectrum with the detection power. An example of measurement
results is shown in Fig. \ref{Rel2} with the curve starting with $P\approx1$. One can see that the relaxation does not follow a single
exponential decay. A bi-exponential fit gives a better result, and provides two time constants: short $T^{'}_{ab}$ and long $T^{''}_{ab}$.

\begin{figure}
\begin{center}
\includegraphics[%
  width=0.95\linewidth,
  keepaspectratio]{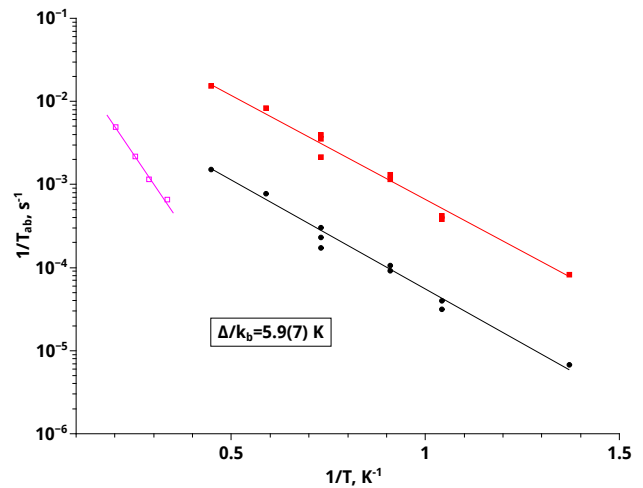}
\end{center}
\caption{(Color online) Nuclear relaxation rates {\color{red}$\blacksquare$} - $1/T^{'}_{ab}$  and \textbf{\textbullet} - $1/T^{''}_{ab}$ as a
function of inverse temperature and exponential fits to the data. Data of van Tol \emph{et al.} \cite{vanTol} are given as
{\color{magenta}$\square$} for comparison.} \label{RelRate}
\end{figure}

The relaxation rates $1/T^{'}_{ab}$ and $1/T^{''}_{ab}$ as functions of the inverse temperature are presented in Fig. \ref{RelRate}. Both
relaxation rates follow an exponential dependence on $\Delta /k_{B}T$ with the activation energy of $\Delta /k_{B}=5.9(7)$ K. We also included
the results of the previous work in strong (8.6 T) magnetic field \cite{vanTol} in the Fig. \ref{RelRate} for comparison.

In a further study of the bi-exponential relaxation, we made measurements in which the nuclear polarization was let to relax from $P_0\approx1$
to a very low value $P\approx 0.03$. Then, for a short time ($30-50$ s) we pumped the \textit{b-c} transition. As a result, the nuclear
polarization was enhanced to $P_{0}\approx0.6$. The subsequent nuclear relaxation measurement revealed that the relaxation was much faster at
the same $P$ value than it was when the measurement was started after the longer pumping with $P_0\approx 1$. A similar result was found after
repeating this measurement for an even smaller starting polarization ($P_{0}\approx 0.3$). The slopes of the relaxation curves were quite the
same at the beginning, meaning that the $T^{'}_{ab}$ does not depend on the starting value of the nuclear polarization.

\section{Discussion}
\subsection{Cross-relaxation rates}
The rate of DNP via the Overhauser effect is defined by the probabilities of the flip-flop and flip-flip transitions $1/T_{ac}$ and $1/T_{bd}$. The flip-flop transition for an isolated donor with isotropic hyperfine interaction is allowed because of the hyperfine mixing of the wave functions. The flip-flop transition relaxation rate in relation to the allowed one is given as $1/T_{ac}=\eta^{2} 1/T_{1e}$ for which
$\eta^{2}=(A/\hbar\gamma_e B_0)^2 \approx 2\times 10^{-7}$ for P donors in 4.6 T magnetic field. This gives $T_{ac}\sim 10^6$ s with measured $T_{1e} \sim 0.2$ s in these experiments. The flip-flip transition, on the other hand, is completely forbidden for the isolated donors and thus the cross-relaxation mechanism for the donor electron with its own nucleus has to involve modulation of the hyperfine interaction by the lattice phonons \cite{Pines57, Jeffries60}. An order
of magnitude evaluation of the relaxation time constant can be done from the following equation \cite{Pines57}
\begin{equation}\label{Tx}
   T_{x}\equiv T_{ac}=\frac{4\pi \hbar^2 \rho}{\gamma_{e}^2 B_{0}^2 kT \gamma^2 A },
\end{equation}
where $\gamma$ is a numerical factor between 10 and 100, and $\rho$ is the density of silicon. For Si:P in 0.3 T (0.8 T) field and temperature of 1.2 K Eq. (\ref{Tx}) gives $T_x=3.4\times 10^{4}$ s ($0.5\times 10^{4}$ s), which is within an order of magnitude agreement with the experimental results of Feher and Gere
\cite{Feher59} $T_{x}\approx 10^5$ s in 0.32 T field and $T_{x}\approx 1.8\times 10^4$ s in 0.8 T field and temperature of 1.25 K. We would like to note also that these two values confirm the $1/B_{0}^2$ dependence of the relaxation time. If we now extrapolate
the data to our field of 4.6 T and temperature of $0.75$ K, we get $T_x\approx 10^3$ s which is in good agreement with our experimental data for
$T^{''}_{ac}\approx1100$ s. So, the phonon induced modulation of the hyperfine interaction\cite{Pines57} can explain the values of the long cross-relaxation time $T^{''}_{ac}$ observed in this work. However, the fast cross-relaxation $T^{'}_{ac}\approx 20$ s in the beginning of the DNP and
the non-zero flip-flip relaxation can not be explained by the theory involved so far.

The observed discrepancy can be explained if we consider the effect of the neighbouring $^{29}$Si on the P cross relaxation. The randomly distributed magnetic nuclei of $^{29}$Si with 4.7$\%$ abundance may create anisotropy in the hyperfine interaction of P with its own nucleus. This leads to a strong enhancement of the forbidden relaxation rates, especially for the donors which have one or more $^{29}$Si in the few closest lattice
shells. For these $^{29}$Si nuclei the anisotropic dipole-dipole interaction with donor electron leads to a fast nuclear relaxation, which scales as $1/r^6$ with the distance between the donor and the $^{29}$Si nucleus \cite{Bloembergen48, Blumberg}. In a similar fashion the proximity of $^{29}$Si may lead to enhancement of the P cross-relaxation. The donors can be separated into two groups: first - having one or several $^{29}$Si in the nearest lattice shell, and second - not having $^{29}$Si nuclei with strong anisotropic interactions
nearby. The first kind of donors have an enhanced cross relaxation and thus the DNP rate, which is seen in the first part of the polarization dynamics of fig. \ref{PolDynamics}. Once these nuclei get fully polarized, the DNP proceeds more slowly with the donors in the second group, and in this case the observed characteristic
time $T^{''}_{ac}$ should be about the same as for the P in isotopically purified $^{28}$Si.

The forbidden cross-relaxation rates also mediates the nuclear relaxation in insulating solids through the nuclear Orbach process\cite{AbragamGoldman}. The temperature dependence of the nuclear relaxation rate is given by the Boltzmann factor:
\begin{equation}\label{Tn_T_dep}
\frac{1}{T_{ab}}= e^{-g_{e}\mu_{B}B/k_{B}T}(\frac{1}{T_{ac}}+\frac{1}{T_{bd}}).
\end{equation}
This means that the nuclear relaxation of the P atoms in the first group should be also faster and could explain our observation of the two time scales
in the nuclear relaxation measurements (figs. \ref{Rel2} and \ref{RelRate}).

\subsection{Nuclear polarization of silicon}
For the normal isotope composition, the $^{29}$Si nuclear spins play an important role in the spin dynamics of shallow donors \cite{Itoh, Dementyev, Ramanathan}. The influence of $^{29}$Si on the P relaxation may complicate further if we consider the possibility of polarizing their spins during the ESR pumping. The polarization transfer between different nuclei is a well known effect.\cite{Solomon1955} The polarization of the $^{29}$Si nuclei in the neighboring lattice sites around the donors should make the P cross relaxation either easier or harder depending on the direction of the polarization. This may lead to enhancement of the cross relaxation transition probabilities in the initial part of the DNP curve (Fig. 3), and make nuclear relaxation faster in the beginning. In other words, the forbidden relaxation time $T^{'}_{ac}$ may depend on the polarization of $^{29}$Si, which is created simultaneously with the OE DNP of phosphorus. This influences to the first fast relaxing group of donors, and should not affect to $T^{''}_{ac}$ and $T^{''}_{ab}$.

\begin{figure}
\begin{center}
\includegraphics[%
  width=0.95\linewidth,
  keepaspectratio]{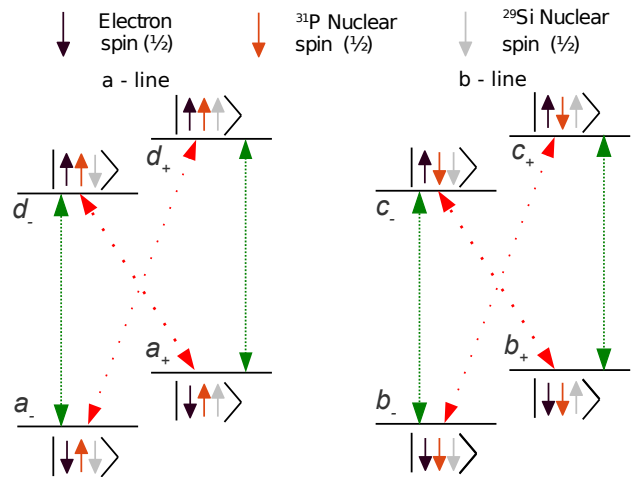}
\end{center}
\caption{(Color online) Hyperfine levels of P donor splitted by the superhyperfine interaction with one $^{29}$Si nucleus. The dashed green arrows indicate the allowed and red dotted arrows forbidden transitions.} \label{diagram}
\end{figure}

ESR technique provides the phosphorus hyperfine state populations, which are proportional to the integrals of the absorption signal. Unfortunately, we could not extract direct information about the $^{29}$Si polarization due to lack of NMR in the experiments. Some information, however, can be extracted from the ESR spectra. If all the $^{29}$Si nuclei are polarized, the superhyperfine interaction shifts the ESR lines in respect to the lines with unpolarized spin states. The sign of the shift depends on the polarization direction of $^{29}$Si. As an example, we consider a splitting of the donor electron spin levels by superhyperfine interaction with one $^{29}$Si nucleus, shown in Fig. \ref{diagram}.
The spin orientations of the $^{29}$Si level scheme are opposite to the P levels due to the negative gyromagnetic factor of $^{29}$Si and the splittings of the allowed transitions are smaller because of the weak superhyperfine interaction ($\lesssim3$ MHz \cite{Feher}). Saturating the whole ESR line of P, $b-c$ for example, induces simultaneously transitions $b_{+}-c_{+}$ and $b_{-}-c_{-}$, which has opposite directions of the $^{29}$Si spins (denoted with $+$ or $-$ subscripts in Fig. \ref{diagram}. The states $b_{+}$ and $c_{-}$ are mixed and they have opposite directions of the electron and $^{29}$Si nuclear spins. The flip-flop transitions between these states have a higher probability than the flip-flip $b_{-}-c_{+}$ transitions. Saturating the ESR line creates OE DNP of the $^{29}$Si in a similar way than in the original Overhauser experiment for nuclei in metals. During the pumping the populations are transferred in both ways: $b_{-}-c_{-}- b_{+}$ and $b_{+}-c_{+}-b_{-}$, and as a result the $^{29}$Si spin populations are redistributed between $b_{+}$ and $b_{-}$ states. Due to the higher rate of the flip-flop transition most of the $^{29}$Si spins end up in the $b_+$ state. Since the resonant field for the $b_{+}-c_{+}$ transition is larger, the ESR line will acquire shift to the right. The magnitude of the shift can be estimated numerically by arranging the $^{29}$Si nuclei randomly into the lattice cites and taking into account known values of hyperfine interactions for each cite \cite{Feher59, Hale69, Ivey75}. For fully polarized sample the evaluation gives $\approx2.8$ G shift.

\begin{figure}
\begin{center}
\includegraphics[%
  width=0.95\linewidth,
  keepaspectratio]{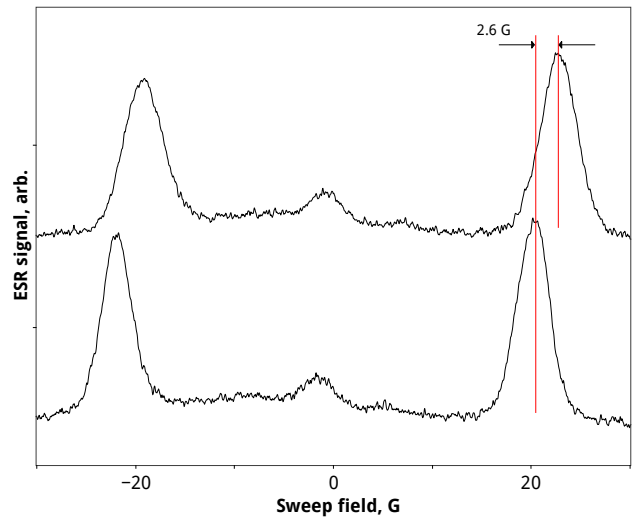}
\end{center}
\caption{Shift of the P ESR lines due to polarization of $^{29}$Si. Upper trace - the sample with maximum polarization. Lower trace - after depolarization by applying FM modulated RF excitation at the frequency of $38.9\pm3$ MHz, corresponding to the NMR transitions of $^{29}$Si in 4.6 T field.}
\label{Shift}
\end{figure}

In the DNP experiments described in Section \ref{sec:Results} the $^{31}$P allowed transitions were pumped for long times, leading to the population transfer from $a$
to $b$ state and vice versa. The relaxation time of the $^{29}$Si nuclear spins is expected to be very long at temperatures of $\sim1$ K. Therefore, it
was difficult to trace the possible shift of the spectrum due to the $^{29}$Si polarization. However, we were able to depolarize the sample by
applying RF excitation at $38.9$ MHz, resonant with the NMR transition of $^{29}$Si, and driving its polarization back to zero. The change in
the spectra resulting from the depolarization is presented in Fig. \ref{Shift}. The upper trace, recorded for the sample with maximum
polarization, is shifted to the right from the unpolarized (lower) trace by $\approx2.6$ G. This, according to the considerations above, proves that
rather high polarization of $^{29}$Si was also reached in this work. 

The rate of $^{29}$Si DNP is the highest for the nuclei closest to the donors. However, due to fast spin diffusion the polarization rapidly propagates into the bulk of the sample  \cite{Bloembergen48}. The rate of the spin diffusion is substantially slower inside the so-called spin diffusion barrier \cite{Bloembergen48, Ramanathan} which is about 10 nm around P donors in our case, and is comparable to the mean distance
between the donors which is about $25 $ nm. After switching off the pumping, the nearest $^{29}$Si will get depolarized by the donors. The
polarization outside the diffusion barrier may decay only because of the nuclear relaxation, which is very slow for remote, non-interacting
$^{29}$Si. Therefore, the $^{29}$Si nuclear polarization may be rather inhomogeneous during the relaxation measurements. However, since the
diffusion barrier is effectively removed during pumping, when the electron spins are depolarized, we expect that the high bulk polarization of
$^{29}$Si has been obtained immediately after the pumping.

\subsection{Efficiency of DNP}
It is well known from the theory of the Overhauser effect \cite{SlichterBook}, that in the high temperature limit ($g_{e}\mu_{B}B\ll k_{B}T$)
the DNP enhancement over the thermal polarization is given by the ratio of the electron and nuclear gyromagnetic ratios $\gamma_{e}/\gamma_{N}$,
which is $\approx$ 1600 for $^{31}$P. However, this is not a correct result in high fields and low temperatures, when the electron polarization
is very high and the nuclear spins are nearly unpolarized. At these conditions a much larger polarization enhancement can be reached
\cite{AbragamGoldman}. In general case the equations for hyperfine state populations due to OE can be found in the text books
\cite{AbragamGoldman}. Here we present the results in the case of long nuclear relaxation time $T_{ab}\gg T_{ac}$ and low temperatures
$g_{e}\mu_{B}B\gg k_{B}T$, which can be obtained from the following considerations.

A microwave excitation saturating fully the electronic \textit{b-c} transition for a long enough ($t_p \gg T_{ac}$) time will establish thermal
equilibrium between $a$ and $c$-states and therefore $n_{b}=n_{c}\approx n_{a}exp(-g_{e}\mu_{B}B/k_{B}T)$. After switching off the pumping, all
the atoms in \emph{c}-state will quickly relax to \emph{b}-state, and we get the relation
\begin{equation}\label{NaNbPol}
   (\frac{n_a}{n_b})_{DNP}=\frac{n_a}{2n_c} =\frac{1}{2}e^{\frac{g_{e}\mu_{B}B}{k_{B}T}}.
\end{equation}
At 4.56 T and 0.2 K this gives a theoretical limit of the Overhauser DNP: $n_{a}/n_{b} \approx 10^{13}$.

So far we have neglected the nuclear relaxation, which will transfer atoms back from $a$ to $b$-state reducing the nuclear polarization. If the
DNP enhancement is balanced by the nuclear relaxation and $n_a \gg n_b$, then from the kinetic equation of \emph{a}-state
\begin{equation}
   \frac{dn_a}{dt}=\frac{n_c}{T_{ac}}-\frac{n_a}{T_{ab}}=0,
\end{equation}
follows that the maximum equilibrium value for the nuclear populations is
\begin{equation}\label{NaNbRel}
   (\frac{n_a}{n_b})_{max}= \frac{n_a}{2n_c}=\frac{T_{ab}}{2T_{ac}}.
\end{equation}

The temperature dependence of our data for the short and long relaxation times in Fig. \ref{RelRate} coincides well with Eq. (\ref{Tn_T_dep}),
with $\Delta/ k_{B}=5.9(7)$ K being close to $g_{e}\mu_{B}B/k_{B} \approx 6.2$ K. Within the temperature range of our current work the forbidden
relaxation time seems to be temperature independent. Therefore from Eqs. \ref{NaNbRel} and \ref{Tn_T_dep} it follows that the maximum
polarization, limited by the nuclear relaxation, has the same temperature dependence as that for the theoretical limit of Overhauser DNP (Eq.
 (\ref{NaNbPol})). From the pre-exponent of $T^{''}_{ab}$ calculated from our data, we found that the maximum polarization is reduced by two orders
of magnitude from the theoretical limit of the Overhauser DNP (Eq. (\ref{NaNbPol})) due to the nuclear relaxation. Including this effect at
$T=$0.2 K we get from Eq. (\ref{NaNbRel}) an estimate of $n_a/ n_b \approx 5\times 10^{11}$. Reaching very high values of the DNP then actually
becomes only a matter of time at these conditions. Pumping for a couple of hours ($t_p\sim 7 T^{''}_{ac}$) will increase $n_a / n_b$ by three
orders of magnitude. The DNP can be done with very low values of the pumping power, enough to fully saturate the allowed ESR transitions,
which is less than 1 $\mu$W in our case.

In summary, from the evolution of the spin polarization during the DNP and the nuclear relaxation measurements, we conclude that the
polarization of neighboring $^{29}$Si spins has an important role in the dynamics of the $^{31}$P nuclear spins. Pumping the allowed ESR
transition lead to the effective DNP of both nuclear ensembles. The populations of the phosphorus hyperfine states were measured directly by
ESR. Record high polarization values were reached after pumping with small excitation power and for relatively short times. We believe that the
possibility of creating ensembles of nuclear spins in ultra-pure spin states demonstrated in this work may find useful applications, e.g. in the
field of quantum computing.

\begin{acknowledgements}
We would like to thank K. Itoh, K. Kono, D. Konstantinov, and A. Tyryshkin for useful discussions. We acknowledge the funding from the Wihuri
Foundation and the Finnish academy grants No. 260531 and 268745.
\end{acknowledgements}

\end{document}